\documentstyle[12pt,equation]{article}
\begin{document}
\newcommand{\tcr}{T_{cr}}
\newcommand{\df}{\delta \phi}
\newcommand{\dkl}{\delta \kappa_{\Lambda}}
\newcommand{\lx}{\lambda}
\newcommand{\Lx}{\Lambda}
\newcommand{\ex}{\epsilon}
\newcommand{\db}{{\bar{\delta}}}
\newcommand{\lb}{{\bar{\lambda}}}
\newcommand{\lt}{\tilde{\lambda}}
\newcommand{\lr}{{\lambda}_R}
\newcommand{\lbr}{{\bar{\lambda}}_R}
\newcommand{\lk}{{\lambda}(k)}
\newcommand{\lbk}{{\bar{\lambda}}(k)}
\newcommand{\ltk}{\tilde{\lambda}(k)}
\newcommand{\mx}{{m}^2}
\newcommand{\mb}{{\bar{m}}^2}
\newcommand{\mt}{\tilde{m}^2}
\newcommand{\mr}{{m}^2_R}
\newcommand{\mk}{{m}^2(k)}
\newcommand{\Pt}{\tilde P}
\newcommand{\Mt}{\tilde M}
\newcommand{\Qt}{\tilde Q}
\newcommand{\Nt}{\tilde N}
\newcommand{\rhb}{\bar{\rho}}
\newcommand{\rht}{\tilde{\rho}}
\newcommand{\rhz}{\rho_0}
\newcommand{\yz}{y_0}
\newcommand{\rhzk}{\rho_0(k)}
\newcommand{\kx}{\kappa}
\newcommand{\kt}{\tilde{\kappa}}
\newcommand{\kk}{\kappa(k)}
\newcommand{\ktk}{\tilde{\kappa}(k)}
\newcommand{\Gammat}{\tilde{\Gamma}}
\newcommand{\Lt}{\tilde{L}}
\newcommand{\Zt}{\tilde{Z}}
\newcommand{\zt}{\tilde{z}}
\newcommand{\zh}{\hat{z}}
\newcommand{\uh}{\hat{u}}
\newcommand{\Mh}{\hat{M}}
\newcommand{\wt}{\tilde{w}}
\newcommand{\etat}{\tilde{\eta}}
\newcommand{\Gammak}{\Gamma_k}
\newcommand{\be}{\begin{equation}}
\newcommand{\ee}{\end{equation}}
\newcommand{\een}{\end{subequations}}
\newcommand{\ben}{\begin{subequations}}
\newcommand{\beq}{\begin{eqalignno}}
\newcommand{\eeq}{\end{eqalignno}}
\pagestyle{empty}
\noindent
OUTP 95-12 P \\
HD-THEP-95-15 \\
CAU-THP-95-10 \\
July 1995
\vspace{2cm}
\begin{center}
{{ \Large  \bf
Solving non-perturbative flow equations
}}\\
\vspace{10mm}
J. Adams$^{\rm a}$, J. Berges$^{\rm b}$,
S. Bornholdt$^{\rm c}$,
F. Freire$^{\rm b}$,
N. Tetradis$^{\rm a}$ and C. Wetterich$^{\rm b}$ \\
\vspace{5mm}
a) Theoretical Physics\\
University of Oxford \\
1 Keble Rd., Oxford OX1 3NP \\
U.K. \\
\vspace{3mm}
b) Institut f\"ur Theoretische Physik \\
Universit\"at Heidelberg \\
Philosophenweg 16, 69120 Heidelberg \\
Germany\\
\vspace{3mm}
c) Institut f\"ur Theoretische Physik \\
Universit\"at Kiel \\
Olshausenstr. 6, 24118 Kiel \\
Germany

\end{center}

\setlength{\baselineskip}{20pt}
\setlength{\textwidth}{13cm}

\vspace{2.cm}
\begin{abstract}
{
Non- perturbative exact flow equations describe the
scale dependence of the effective average action.
We present a numerical solution for an approximate form
of the flow equation for the potential in a three-dimensional
$N$-component scalar field theory. The critical behaviour,
with associated critical exponents, can be inferred with
good accuracy.
}
\end{abstract}
\clearpage
\setlength{\baselineskip}{15pt}
\setlength{\textwidth}{16cm}
\pagestyle{plain}
\setcounter{page}{1}

\newpage

\setcounter{equation}{0}

Exact non-perturbative renormalization group equations
describing the scale dependence of some type of effective
action have been known for a long time \cite{rengr}. They account for
the consecutive inclusion of fluctuations in a field theory.
There exist many versions of such equations which are, if
exact, all equivalent, since they all describe in one way
or another properties of the (Euclidean) functional integral
which defines the theory. The most difficult part, however, is not
so much to derive an exact flow equation - this usually
follows from simple manipulations of the functional integral.
The challenge is rather to find a formulation which can be used
for practical computations beyond perturbation theory. For such
practical purposes the most general form of the effective action
has always to be truncated. Solutions of the truncated equations
are only approximations to the exact flow equations. If a small
parameter is available one can often organize a series of truncations
as a power series in the small parameter - this is how
usual perturbation theory is recovered. If no small parameter
is known a priori one has to use as much as possible the knowledge
of properties of the model in order to conceive a useful
truncation. This is where it becomes important what effective
action is selected and what the particular form of the flow
equation is. Only if the effective action has a simple physical
meaning so that its couplings can also be understood by
other methods, and if the evolution equation has a form which
incorporates directly the most prominent properties of the system,
there is a realistic chance of getting a working non-perturbative
tool. Beyond perturbation theory exact flow equations should,
therefore, not be viewed as mainly a consistent (but unsolvable)
mathematical system, but rather as a convenient
starting point for appropriate non-perturbative expansion methods.

Typically, the truncated flow equations constitute a system
of non-linear partial differential equations which can be
solved by numerical methods. It is the aim of this letter
to report on the development of algorithms which are
adapted to the specific numerical problems of this approach.
As an example we investigate here the $N$-component Heisenberg
models in three dimensions. This is a classical testing ground
for non-perturbative methods. We will extract the critical
exponents of these models from a numerical solution of the flow
equation for the scale dependent effective potential, as obtained
from the lowest order in a
derivative expansion of the effective action.

Recently the average action $\Gamma_k$
\cite{average} has been proposed
as the free energy with an infrared cutoff $\sim k$. It is
formulated in continuous space so that all symmetries
of the model are preserved. More precisely the
average action includes the effects
of all fluctuations with momenta $q^2 > k^2$. In the limit
$k \rightarrow 0$ it becomes the standard effective action
(the generating functional of the 1PI Green functions), while for
$k \rightarrow \infty$ it equals the classical action.
There is a simple functional integral representation \cite{average}
of $\Gamma_k$ also for $k > 0$ such that its couplings can,
in principle, also be estimated by alternative methods.
The average action is a coarse-grained free energy
in the sense that short distance fluctuations are already
integrated out.
\footnote{
One should not use $\Gamma_k$ directly to replace the
action in the functional integral, as this is the role
of the ``cutoff action'' used in earlier approaches
\cite{rengr}. For this purpose an explicit ultraviolet cutoff
term has to be added in order to prevent double counting
of the short distance modes. This cutoff term is
easily computed in simple theories of
scalars and fermions \cite{average}, much more involved for an
abelian gauge theory \cite{gauge}, and not yet found for
non-abelian gauge theories.}
The exact non-perturbative flow equation for $\Gamma_k$
takes the simple form of a renormalization group improved
one-loop equation \cite{exact}
\be
\partial_t \Gamma_k = k \frac{\partial}{\partial k} \Gamma_k
= \frac{1}{2} {\rm Tr} \tilde{\partial}_t
\ln \left( \Gamma_k^{(2)} + R_k \right),
\label{one} \ee
where $\tilde{\partial}_t$ acts only on the infrared cutoff
piece $R_k$ as
\be
\tilde{\partial}_t \ln \left( \Gamma_k^{(2)} + R_k \right)
= \left( \Gamma_k^{(2)} + R_k \right)^{-1}
 k \frac{\partial}{\partial k} R_k.
\label{two} \ee
The trace involves a momentum integration and summation over
internal indices. Most importantly, the relevant infrared
properties appear directly in the form of the exact inverse
average propagator $\Gamma_k^{(2)}$, which is the matrix of
second functional derivatives with respect to the fields.
There is always only one momentum integration - multi-loops
are not needed - which is, for suitable $R_k$, both infrared and
ultraviolet finite.

The flow equation (\ref{one}) seems to be a good starting point
for a non-perturbative approach. Nevertheless,
it remains a complicated functional differential equation without
any chance to be solved exactly. Approximate solutions need
truncations - and the crux of the problem lies there.
One possibility is to keep only a few invariants in $\Gamma_k$
and, thereby, reduce eq. (\ref{one}) to a finite set of ordinary
differential equations for a finite number of couplings. Very
satisfactory results have already been obtained this way
\cite{expon}. Much more information is contained if, instead of a
finite number of couplings, arbitrary functions of one or
several parameters are considered.
Examples are an arbitrary field dependence of the average potential
$U_k(\rho)$ in scalar theories (with $\rho = \frac{1}{2} \phi^2$)
\cite{expon,morris},
or an arbitrary momentum dependence of the two-point
function $G(p)$ \cite{papen} or even the four-point function
$\lx(q_1,q_2,q_3,q_4)$ \cite{ellwanger}. The functional
differential equation becomes then a partial differential
equation for a function of at least two variables,
e.g. $U(\rho,t)=U_k(\rho)$, or a system
of partial differential equations for several such functions.
If one does not want to resort to further approximations
at this level, one needs appropriate tools for
solving this type of partial differential equations.
Analytical solutions can be found only in certain limiting
cases and for most purposes numerical solutions seem the
adequate tool. This is not so straightforward as it may
seem at first sight, since the solutions of the differential equations
have in general a highly unstable character
due to the presence of relevant
parameters. Fine tuning of initial conditions is needed
in order to be near a phase transition - or, in
particle physics language, to have renormalized masses much
smaller than the ultraviolet cutoff. The algorithm for a numerical
solution must be compatible with this situation and guarantee
numerical stability of the critical solution which corresponds
to the phase transition.

In this letter we demonstrate the
capacities of such a non-perturbative method by computing the
$k$-dependent average potential $U_k(\rho)$
for an $N$-component scalar field
theory directly in three dimensions. For
$k \rightarrow 0$ this gives the free energy which encodes the
equation of state of the system. In the phase with spontaneous
symmetry breaking the minimum of the potential occurs for $k=0$ at
$\rhz \not= 0$. The massless Goldstone excitations around this
minimum are notoriously difficult to treat by alternative
methods. For example, the standard loop expansion is order
by order highly infrared divergent. In the symmetric phase
the minimum of $U_k(\rho)$ ends at $\rhz=0$ for $k=0$.
The two phases are separated by a scaling solution for which
$U_k/k^3$ becomes independent of $k$ once expressed in terms
of a suitably rescaled field variable $\rht$.

Our truncation is the lowest order in a systematic derivative
expansion of $\Gamma_k$ \cite{average,expon,morris}
\be
\Gamma_k = \int d^dx \bigl\{
U_k(\rho) + \frac{1}{2} Z_k \partial^{\mu} \phi_a
\partial_{\mu} \phi^a \bigr\}.
\label{three} \ee
Here $\phi^a$ denotes the $N$-component real scalar field and
$\rho = \frac{1}{2} \phi^a \phi_a$. We keep for the
potential term the most general $O(N)$-symmetric
form $U_k(\rho)$, whereas the wave function renormalization
is approximated by one $k$-dependent parameter.
Next order in the derivative expansion would be the
generalization to a $\rho$-dependent wavefunction
renormalization $Z_k(\rho)$ plus a function
$Y_k(\rho)$ accounting for a possible different
index structure of the kinetic term for $N \geq 2$
\cite{average,expon}.
Going further would require the consideration of
terms with four derivatives and so on.
For the three-dimensional scalar theory the anomalous dimension
$\eta$ is known to be small and the derivative expansion
is, therefore, expected to give a reliable approximation \cite{expon}.
The main reason is that for $\eta=0$ the kinetic term in the
$k$-dependent inverse propagator must be exactly proportional to
$q^2$ both for $q^2 \rightarrow 0$ and
$q^2 \rightarrow \infty$. This holds for arbitrary constant
``background'' field $\phi^a$. Similar, although less
stringent, arguments concern the smallness of the
$\rho$-dependence of the kinetic term \cite{expon}. For the
scaling solution for $N=1$ this weak $\rho$-dependence has
been established explicitly \cite{morris}.
We finally mention that $\eta$ is proportional to a small parameter
$\lx/8 \pi^2$, where $\lx$ is a suitably defined quartic scalar
coupling. (For the scaling solution $\lx$ takes a fixed point
value $\lx_*/8 \pi^2 = 0.12$ for $N=1$.)
We expect that the derivative expansion can be understood
as an expansion in this small parameter. However, since
some of the other parameters effectively
behave $\sim \lx^{-1}$ this
expansion is not equivalent to the usual perturbative
polynomial series in $\lx$.

For a study of the behaviour in the vicinity of the phase transition
it is convenient to work
with dimensionless renormalized fields
\footnote{We keep the number of dimensions $d$ arbitrary and specialize
only later to $d=3$.}
\beq
\rht =& Z_k k^{2-d} \rho \nonumber \\
u_k(\rht) =& k^{-d} U_k(\rho).
\label{four} \eeq
With the truncation of eq. (\ref{three}) the exact
evolution equation for $u'_k \equiv \partial u_k/\partial \rht$
\cite{expon}
reduces then to the partial differential equation
\beq
\frac{\partial u'_k}{\partial t} =~&(-2 + \eta) u'_k +(d-2+ \eta)\rht u_k''
\nonumber \\
&- 2 v_d (N-1) u_k'' l^d_1(u_k';\eta)
- 2 v_d (3 u_k'' + 2 \rht u_k''') l^d_1(u_k'+2 \rht u_k'';\eta),
\label{five}
\eeq
where $t = \ln \left( k/\Lx \right)$, with $\Lx$
the ultraviolet cutoff of the theory.
The anomalous dimension $\eta$ is defined by
\be
\eta = - \frac{\partial}{\partial t} \ln Z_k
\label{six} \ee
and
\be
v_d^{-1} = 2^{d+1} \pi^{\frac{d}{2}} \Gamma \left( \frac{d}{2}
\right),
\label{seven} \ee
with $v_3 = 1/8 \pi^2$.
The ``threshold'' functions $l^d_n(w;\eta)$
result from the momentum integration on the
r.h.s. of eq. (\ref{one}), and read for $n \geq 1$, with
$y=q^2/k^2$
\beq
l^d_n(w;\eta) =
{}~&- n \int^{\infty}_0 dy y^{\frac{d}{2}+1}
\frac{\partial r(y)}{\partial y}
\left[ y(1+r(y)) + w \right]^{-(n+1)}
\nonumber \\
&- \frac{n}{2} \eta \int^{\infty}_0 dy y^{\frac{d}{2}}
r(y)
\left[ y(1+r(y)) + w \right]^{-(n+1)}.
\label{eight} \eeq
Here $r(y)$ depends on the choice of the momentum dependence
of the infrared cutoff and we employ
\be
r(y) = \frac{e^{-y}}{1 - e^{-y}}.
\label{nine} \ee
This choice has the property
$\lim_{q^2 \rightarrow 0} R_k/Z_k k^2 =
\lim_{y \rightarrow 0} y r(y) = 1$, whereas for $q^2 \gg k^2$
the effect of the infrared cutoff is exponentially suppressed.
The ``threshold'' functions account for the decoupling
of modes with mass larger than $k$ and decrease rapidly for
$w \gg 1$. For our purpose we use numerical fits for these
functions. Finally, the anomalous dimension is given
in our truncation by
\be
\eta(k) =  \frac{16 v_d}{d} \kx \lx^{2} m^d_{2,2}(2 \lx \kx),
\label{ten} \ee
with $\kx$ the location of the minimum of the potential
and $\lx$ the quartic coupling
\beq
u'_k(\kx)=&0 \nonumber \\
u''_k(\kx)=&\lx.
\label{eleven} \eeq
The function $m^d_{2,2}$ is given by \cite{expon}
\beq
m^d_{2,2}(w) = & \int_0^{\infty} dy y^{\frac{d}{2}-2}
\frac{1 + r +y \frac{\partial r}{ \partial y}}{(1+r)^2
\left[(1+r)y +w \right]^2 }
\nonumber \\
&\biggl\{
2 y \frac{\partial r}{ \partial y}
+ 2 \left(y \frac{\partial }{ \partial y} \right)^2 r
- 2 y^2 \left( 1 +r + y \frac{\partial r}{ \partial y} \right)
\frac{\partial r}{ \partial y}
\left[ \frac{1}{(1+r)y} + \frac{1}{(1+r)y + w} \right]
\biggr\}.
\nonumber \\
{}~&~
\label{twelve} \eeq
We point out that the argument
$2 \lx \kx$ turns out generically to
be of order one for the scaling solution. Therefore,
$\kx \sim \lx^{-1}$ and the mass effects are important,
in contrast to perturbation theory where they are
treated as small quantities $\sim \lx$.

Our aim is the development of algorithms for the numerical
solution of the partial differential equation
(\ref{five}), and a comparison with previously used expansion
methods. Near the phase transition the
trajectory spends most of the ``time'' $t$ in the vicinity of
the critical $k$-independent scaling solution given by
$\partial_t u'_* (\rht) = 0$.
\footnote{The resulting ordinary differential
equation for $u_*(\rht)$ has already been solved numerically
for a somewhat different choice of the infrared cutoff
\cite{morris}.}
Only at the end of the running the ``near-critical''
trajectories deviate from the scaling solution. For
$k \rightarrow 0$ they either end up in the symmetric phase with
$\kx =0$ and positive constant mass term $m^2$ such that
$u'_k(0) \sim m^2/k^2$; or they lead to a non-vanishing
constant $\rhz$ indicating spontaneous symmetry breaking with
$\kx \rightarrow Z_0 k^{2-d} \rhz$. The equation of state
involves the potential $U_0(\rho)$ for
temperatures away from the critical temperature.
Its computation requires the
solution for the running
away from the critical trajectory which involves the
full partial differential equation (\ref{five}).
We have developed two alternative numerical approaches
which we briefly describe in the following.
For both methods we replace the variable $\rht$ by
a discrete set of points $\rht_i$, $i=1,...,m$.

I) Consider first what happens if we make a Taylor
expansion of $u_k$ around some arbitrary point
$\rht_i$
\be
u_k(\rht) = \sum_{n=0}^{\infty}
\frac{1}{n!} u^{(n)}_i
(\rht-\rht_i)^n,
\label{thirteen} \ee
with $u^{(n)}_i(k)=u^{(n)}_k(\rht_i)$.
The potential is then
described by infinitely many couplings
$u^{(n)}_i(k)$.
The flow equations for these couplings are obtained from
appropriate $\rht$-derivatives of eq. (\ref{five}) evaluated
at $\rht=\rht_i$. We observe that the
flow equation for $u^{(1)}_i$ involves
$u^{(1)}_i$, $u^{(2)}_i$ and $u^{(3)}_i$, the one for
$u^{(2)}_i$ needs in addition $u^{(4)}_i$, and the system
is never closed. Our first approach
considers at every point $\rht_i$ the differential
equations for $u^{(1)}_i$ and $u^{(2)}_i$. The
``missing'' couplings $u^{(3)}_i$ and $u^{(4)}_i$
appearing in these equations are determined by
matching the expansion around
$\rht_i$ with similar expansions around different
points $\rht_{j \not= i}$. More precisely, the
matching is done by equating fourth order polynomial expansions
of $u'_k(\rht)$ around two neighbouring points $\rht_i$ and
$\rht_{i+1}$ at half-distance, and similarly for $u''_k(\rht)$
\beq
&(u'_k)_i \left( \frac{\rht_i + \rht_{i+1}}{2} \right)
\equiv u^{(1)}_i + u^{(2)}_i \frac{\rht_{i+1} - \rht_i}{2}
+ u^{(3)}_i \frac{(\rht_{i+1} - \rht_i)^2}{8}
+ u^{(4)}_i \frac{(\rht_{i+1} - \rht_i)^3}{48}
\nonumber \\
=~ &(u'_k)_{i+1} \left( \frac{\rht_i + \rht_{i+1}}{2} \right)
\equiv u^{(1)}_{i+1} - u^{(2)}_{i+1} \frac{\rht_{i+1} - \rht_i}{2}
+ u^{(3)}_{i+1} \frac{(\rht_{i+1} - \rht_i)^2}{8}
- u^{(4)}_{i+1} \frac{(\rht_{i+1} - \rht_i)^3}{48}
\nonumber \\
{}~&~
\label{fourteen} \\
&(u''_k)_i \left( \frac{\rht_i + \rht_{i+1}}{2} \right)
\equiv u^{(2)}_i + u^{(3)}_i \frac{\rht_{i+1} - \rht_i}{2}
+ u^{(4)}_i \frac{(\rht_{i+1} - \rht_i)^2}{8}
\nonumber \\
=~ &(u''_k)_{i+1} \left( \frac{\rht_i + \rht_{i+1}}{2} \right)
\equiv u^{(2)}_{i+1} - u^{(3)}_{i+1} \frac{\rht_{i+1} - \rht_i}{2}
+ u^{(4)}_{i+1} \frac{(\rht_{i+1} - \rht_i)^2}{8}.
\label{fifteen} \eeq
Combining eqs. (\ref{fourteen}) and (\ref{fifteen}) at all
$m-1$ intermediate points ($i=1,...,m$) gives an algebraic system
of $2m-2$ equations for the $2m$ unknowns $u^{(3)}_i$
and $u^{(4)}_i$.
In order to obtain the remaining two necessary equations,
we also match for the initial and end points
the third derivative
of the expansions, at $(\rht_1+\rht_2)/2$ and
$(\rht_{m-1}+\rht_m)/2$ respectively.
Together these equations make up an
algebraic system which has a unique solution. This allows
one to express the couplings $u^{(3)}_i$ and $u^{(4)}_i$
as functions of the couplings $u^{(1)}_j$
and $u^{(2)}_j$, $j=1,...,m$.
The integration of the remaining system of $2m$ differential
equations for the couplings $u^{(1)}_i$, $u^{(2)}_i$
is done with the fifth order Runge-Kutta algorithm using
the embedded fourth order method for precision control.
The function $u'_k(\rht)$ is finally reconstructed by patching
the fourth order Taylor expansions around $\rht_i$ together
at half-distance around neighbouring points.
The polynomial patching improves with decreasing distance between
neighbouring expansion points. This can be used to check the
stability of numerical results.

We note that the matching conditions of eqs. (\ref{fourteen}),
(\ref{fifteen}) guarantee continuity of
$u'_k(\rht)$ and $u''_k(\rht)$. (Further smoothening
could be applied for $u^{(3)}_k(\rht)$ if needed.)
Furthermore, they imply that the differential equations
for $u^{(1)}_i$ and $u^{(2)}_i$ do not only incorporate
information from directly neighbouring points, but from
the whole range of points, as implied by the
algebraic solution for $u^{(3)}_i$ and $u^{(4)}_i$. Nevertheless,
one observes that the contributions of
$u^{(1,2)}_j$
to $u^{(3,4)}_i$  $(j\ne i)$ rapidly decrease with increasing $|i-j|$.
The decoupling from distant points could be used to
obtain an approximate expression for $u^{(3,4)}_i$
which becomes useful if a large number of expansion points is
considered. For the computation of the critical exponents
we expand here the potential around 10 points.
As long as the minimum of the potential is located
away from the origin we
choose expansion points $\rht_i$ proportional to $\kx$.
If the minimum is at or very close to
the origin we use instead expansion points corresponding to
fixed values of $\rho$ instead of fixed $\rht$.

II) For the alternative approach we first consider a standard discretized
version of eq. (\ref{five})
\be
k \frac{u^{(1)}_i(k+\Delta k)-u^{(1)}_i(k)}{\Delta k}
= F \left[ u^{(1)}_i(k), u^{(2)}_i(k), u^{(3)}_i(k) \right].
\label{sixteen} \ee
The higher $\rht$-derivatives
$u^{(2)}_i(k), u^{(3)}_i(k)$ can be inferred from the differences of
the values
of $u^{(1)}_{i \pm 1}(k)$ and the integration seems
straightforward. The main problem with this approach is
the appearance of numerical instabilities, with the numerical
solution becoming strongly oscillating after a few integration
steps. These instabilites are suppressed if the r.h.s. of
eq. (\ref{sixteen}) is evaluated at
$k+\Delta k$ instead of $k$. This leads us to replace the
partial differential equation (\ref{five}) by the system
of $m$ algebraic equations
\be
k \frac{u^{(1)}_i(k+\Delta k)-u^{(1)}_i(k)}{\Delta k}
= F \left[ u^{(1)}_i(k+\Delta k),
u^{(2)}_i(k+\Delta k), u^{(3)}_i(k+\Delta k) \right],
\label{seventeen} \ee
with the higher $\rht$-derivatives expressed in terms of
$u^{(1)}_{i \pm 1}(k+\Delta k)$.
\footnote{ For the end points $\rht_1$, $\rht_m$, the
higher derivatives can be obtained at the same level
of accuracy in terms of $u^{(1)}_{1,2,3,4}$ and
$u^{(1)}_{m,m-1,m-2,m-3}$ respectively.}
At every step the $m$ unknowns
$u^{(1)}_{i}(k+\Delta k)$
can be calculated with the Newton-Raphson method.
Further improvement in the accuracy can be achieved by
making use of the solution at $k$ and $k-\Delta k$ for the
calculation of the solution at $k + \Delta k$ according to
\beq
k \frac{3 u^{(1)}_i(k+\Delta k)-4 u^{(1)}_i(k)
+ u^{(1)}_i(k-\Delta k)}{2 \Delta k}
=& F \left[ u^{(1)}_i(k+\Delta k),
u^{(2)}_i(k+\Delta k), u^{(3)}_i(k+\Delta k) \right].
\nonumber \\
{}~&~
\label{extra} \eeq
The expression on the l.h.s. approximates the
$k$-derivative at $k+\Delta k$ with an accuracy of
{\cal O}($|\Delta k|^2$).
We use 60 points for the discretization of the variable
$\rht$ and a varying number of $k$ steps (around 500) until
stability of the results is obtained.

The use of two algorithms for the integration of
eq. (\ref{five}) provides a
good check for possible systematic numerical
uncertainties. The two methods give results which
agree at the 0.3 \% level and the difference
is most likely due to the different use of fits for the
``threshold'' functions. We expect
the numerical solution to be an approximation of the solution
of the partial differential equation (\ref{five}) with the same
level of accuracy. This has to be compared with the uncertainty
induced by the omission of the higher derivative terms
in the average action. The latter is expected to
be of the order of $\eta$, as we discussed earlier,
and is the main source of
error for the results presented in the following.

In fig. 1 we present the results of the numerical integration
of eq. (\ref{five}) for $d=3$ and $N=1$.
The function $u'_k(\rht)$ is plotted
for various values of $t= \ln(k/\Lx)$. The evolution starts
at $k=\Lx$ ($t=0$)  where the average potential is equal to
the classical potential (no effective integration of modes
has been performed). We start with a quartic classical potential
parametrized as
\be
u'_{\Lx}(\rht) = \lx_\Lx (\rht - \kx_\Lx).
\label{eighteen} \ee
We arbitrarily choose $\lx_\Lx=0.1$ and fine tune $\kx_\Lx$
so that a scaling solution is approached at later stages
of the evolution. There is a critical value
$\kx_{cr} \simeq 6.396 \times 10^{-2}$
for which the evolution leads to the scaling solution
without ever deviating from it.
For the results in fig. 1 a value
$\kx_\Lx$ slightly smaller than $\kx_{cr}$
is used. As $k$ is lowered
(and $t$ turns negative), $u'_k(\rht)$ deviates from its initial
linear shape. Subsequently it evolves towards a form which is
independent of $k$ and corresponds to the scaling
solution $\partial_t u'_* (\rht) = 0$. It spends a long ``time''
$t$ - which can be rendered arbitrarily long through appropriate
fine tuning of $\kx_\Lx$ -
in the vicinity of the scaling solution. During this
``time'', the minimum of the potential $u'_k(\rht)$
takes a fixed value $\kx_*$,
while the minimum of $U_k(\rho)$ evolves towards zero according to
\be
\rhz(k) = k \kx_* / Z_k.
\label{nineteen} \ee
The longer $u'_k(\rht)$ stays near the scaling solution, the
smaller the resulting value of $\rhz(k)$ when the
system deviates from it.
As this value determines the mass scale for the
renormalized theory at $k=0$, the
scaling solution governs the behaviour of the system
very close to the phase transition, where the characteristic
mass scale goes to zero.
Another important property of the ``near-critical''
trajectories, which spend a long ``time'' $t$ near
the scaling solution, is that they become insensitive
to the details of the classical theory which determine the
initial conditions for the evolution. After $u'_k(\rht)$ has
evolved away from its scaling form $u'_*(\rht)$, its
shape is independent of the choice of $\lx_\Lx$ for
the classical theory.
This property gives rise to the universal critical behaviour
near second order phase transitions.
For the solution depicted in
fig. 1 $u_k(\rht)$ evolves in such a way that its minimum
runs to zero with $u'_k(0)$ subsequently increasing.
Eventually the theory settles down in the
symmetric phase with a positive constant renormalized mass term
$m^2 = k^2 u'_k(0) $ as $k \rightarrow 0$.
Another possibility is that the system ends up in the
phase with spontaneous symmetry breaking. In this case
$\kx$ grows in such a way that
$\rhz(k)$ approaches a constant value
for $k \rightarrow 0$.

The approach to the scaling solution and the deviation
from it can also be seen in fig. 2. The evolution of the
running parameters $\kx(t)$, $\lx(t)$ starts
with their initial classical values, leads to
fixed point values $\kx_*$, $\lx_*$ near the scaling solution,
and finally ends up in the symmetric phase
($\kx$ runs to zero).
Similarly the anomalous dimension $\eta(k)$, which is given
by eq. (\ref{ten}), takes a fixed point value $\eta_*$
when the scaling solution is approached.
During this part of the evolution the wave function
renormalization is given by
\be
Z_k \sim k^{-\eta_*}
\label{twenty} \ee
according to eq. (\ref{six}). When the parts of the evolution towards
and away from the fixed point become negligible
compared to the evolution near the fixed point
- that is, very close to the phase transition -
eq. (\ref{twenty}) becomes a very good approximation for
sufficiently low $k$.
This indicates that $\eta_*$ can be identified with the
critical exponent $\eta$.
For the solution of fig. 2 ($N=1$) we find
$\kx_*=4.07 \times 10^{-2}$, $\lx_*=9.04$ and
$\eta_*=4.4 \times 10^{-2}$.

As we have already mentioned the details of
the renormalized theory in the vicinity of the phase
transition are independent of the classical coupling
$\lx_\Lx$. Moreover, the critical theory can
be parametrized in terms of critical exponents
\cite{fisher}, an example of
which is the anomalous dimension $\eta$.
These exponents are universal quantities which depend
only on the dimensionality of the system and its internal
symmetries. For our three-dimensional theory they depend only on the
value of $N$ and
can be easily extracted from our results. We concentrate on
the exponent $\nu$, which parametrizes the behaviour of the
renormalized mass in the critical region. The other exponents
are not independent quantities, but
can be determined from $\eta$ and $\nu$ through universal
scaling laws \cite{fisher}. We define the exponent $\nu$ through
the renormalized mass term in the symmetric phase
\be
m^2 =  \frac{1}{Z_k} \frac{d U_k(0)}{d\rho}
= k^2 u'_k(0)~~~~~~~{\rm for}~~k \rightarrow 0.
\label{twentyone} \ee
The behaviour of $m^2$ in the critical region
depends only on the distance from the phase transition, which
can be expressed in terms of the difference of
$\kx_\Lx$ from the critical value $\kx_{cr}$ for which
the renormalized theory has exactly $m^2 =0$.
The exponent $\nu$ is determined from the relation
\be
m^2 \sim |\dkl|^{2 \nu} = |\kx_\Lx - \kx_{cr}|^{2 \nu}.
\label{twentytwo} \ee
For a determination of $\nu$ from our results we calculate
$m^2$ for various values of $\kx_\Lx$ near $\kx_{cr}$.
We subsequently plot $\ln(m^2)$ as a function
of $\ln|\dkl|$. This curve becomes linear for
$\dkl \rightarrow 0$ and we
obtain $\nu$ from the constant slope.
In the past the critical exponents of the $O(N)$-symmetric
theory were calculated from truncated versions of
the partial differential equation (\ref{five}) \cite{expon}.
The strategy was to turn eq. (\ref{five}) into an infinite
system of ordinary differential equations for the
coefficients of a Taylor expansion analogous to eq. (\ref{thirteen})
around the ``running'' minimum of the potential.
This infinite system was approximately
solved by neglecting $\rht$-derivatives
of $u_k(\rht)$ higher
than a given order. The apparent convergence of the procedure
was checked by enlarging the level of truncation.
We now have an alternative way of estimating the accuracy of this
method. Our numerical solution of the partial differential
equation (\ref{five}) corresponds to an infinite level of
truncation where all the higher derivatives are taken into
account.
In table 1 we present results obtained through the procedure
of successive truncations and through our numerical solution
for $N=3$. We give the values of $\kx$, $\lx$, $u^{(3)}_k(\kx)$
for the scaling solution and the critical exponents
$\eta$, $\nu$. We observe how the results stabilize as more
$\rht$-derivatives of $u_k(\rht)$ at $\rht=\kx$ and the
anomalous dimension are taken into account. The last line
gives the results of our numerical solution of eq. (\ref{five}).
By comparing with the previous line we conclude
that the inclusion of all the $\rht$-derivatives higher than
$u_k^{(6)}(\kx)$ and the term $\sim \eta$ in the
``threshold'' function of eq. (\ref{eight})
generates an improvement of less than
1 \% for the results. This is a lot smaller than the error
induced by the omission of the higher derivative terms
in the average action, which typically generates an uncertainty
of the order of the anomalous dimension.
In table 2 we compare our values for the critical exponents
with more accurate results obtained with other methods
(such as the $\epsilon$-expansion, summed perturbation theory
at fixed dimension, lattice calculations and the $1/N$-expansion).
As expected $\eta$ is rather poorly determined since it is
the quantity most seriously affected by the omission of the
higher derivative terms in the average action. The exponent
$\nu$ is in agreement with the known results at the
1-5 \% level, with a discrepancy roughly equal to the
value of $\eta$ for various $N$.

In conclusion, the shape of the average potential is under
good quantitative control for every scale $k$. This
permits a quantitative understanding of the most
important properties of the system at every length
scale. Our investigation is not restricted to the
behaviour near the phase transition on which we have
concentrated here because it is
the most difficult to handle numerically. Also the
initial form of the potential does not have
to be of the quartic form of eq. (\ref{eighteen}). Arbitrary
general short-distance potential can be studied. For example,
the tricritical point for the transition to a
first order phase transition can be investigated with
our numerical methods. Furthermore, the flow
equation is well defined for arbitrary continuous $d$.
This will permit an explicit check of the
validity of the $\epsilon$-expansion for more detailed
quantities characterizing the equation of state.

\newpage

\section*{Tables}

\begin{table} [h]
\renewcommand{\arraystretch}{1.5}
\hspace*{\fill}
\begin{tabular}{|c|c|c|c|c|c|}	\hline

& $\kx_*$
& $\lx_*$
& $u^{(3)}_*$
& $\eta$
& $\nu$
\\ \hline \hline
a
& $6.57 \times 10^{-2}$
& 11.5
&
&
& 0.745
\\ \hline
b
& $8.01 \times 10^{-2}$
& 7.27
& 52.8
&
& 0.794
\\ \hline
c
& $7.86 \times 10^{-2}$
& 6.64
& 42.0
& $3.6 \times 10^{-2}$
& 0.760
\\ \hline
d
& $7.75 \times 10^{-2}$
& 6.94
& 43.5
& $3.8 \times 10^{-2}$
& 0.753
\\ \hline
e
& $7.71 \times 10^{-2}$
& 7.03
& 43.4
& $3.8 \times 10^{-2}$
& 0.752
\\ \hline
f
& $7.64 \times 10^{-2}$
& 7.07
& 44.2
& $3.8 \times 10^{-2}$
& 0.747
\\ \hline
\end{tabular}
\hspace*{\fill}
\renewcommand{\arraystretch}{1}
\caption[y]
{
The minimum $\kx$ of the
potential $u_k(\rht)$, the derivatives
$\lx=u''(\kx)$, $u_k^{(3)}(\kx)$
for the scaling solution, and
the critical exponents $\eta$ and $\nu$,
in various approximations: (a)-(e) from
ref. \cite{expon}
and (f) from the present
letter. $N=3$. \\
a) Truncation where only the evolution of $\kx$ and $\lx$ is
considered and
higher derivatives of the potential and the anomalous
dimension are neglected. \\
b) $\kx$, $\lx$, $u^{(3)}_k(\kx)$ are included. \\
c) $\kx$, $\lx$, $u^{(3)}_k(\kx)$ are included and $\eta$ is approximated
by eq. (\ref{ten}). \\
d) with five parameters: $\kx$, $\lx$, $u^{(3)}_k(\kx)$, $u^{(4)}_k(\kx)$
and $\eta$. \\
e) as in d) and in addition
$u^{(5)}_k(\kx)$, $u^{(6)}_k(\kx)$ are estimated. \\
f) The partial differential equation (\ref{five})
for $u'_k(\rht)$ is solved numerically
and $\eta$ is approximated by eq. (\ref{ten}). \\
}
\end{table}

\newpage

\begin{table} [h]
\renewcommand{\arraystretch}{1.5}
\hspace*{\fill}
\begin{tabular}{|c||c|c||c|c|}
\hline
$N$
&\multicolumn{2}{c||}{$\nu$}
&\multicolumn{2}{c|}{$\eta$}
\\
\hline \hline

&
&$0.6300(15)^a$
&
&$0.032(3)^{a}$
\\
1
&0.643
&$0.6310(15)^{b}$
&0.044
&$0.0375(25)^{b}$
\\

&
&$0.6305(15)^{c}$
&
&
\\ \hline

&
&$0.6695(20)^{a}$
&
&$0.033(4)^{a}$
\\
2
&0.697
&$0.671(5)^{b}$
&0.042
&$0.040(3)^{b}$
\\

&
&$0.672(7)^{c}$
&
&
\\ \hline

&
&$0.705(3)^{a}$
&
&$0.033(4)^{a}$
\\
3
&0.747
&$0.710(7)^{b}$
&0.038
&$0.040(3)^{b}$
\\

&
&$0.715(20)^{c}$
&
&
\\ \hline
4
&0.787
&
&0.034
&
\\  \hline
10
&0.904
&$0.877^{d}$
&0.019
&$0.025^{d}$
\\ \hline
100
&0.990
&$0.989^{d}$
&0.002
&$0.003^{d}$
\\ \hline
\end{tabular}
\hspace*{\fill}
\renewcommand{\arraystretch}{1}
\caption[y]
{
Critical exponents $\nu$ and $\eta$
for various values of $N$.
For comparison we list results obtained with other methods as summarized in
\cite{zinn} and \cite{journal}: \\
a) From summed perturbation series in fixed dimension 3 at six-loop order. \\
b) From the $\ex$-expansion at order $\ex^5$. \\
c) From lattice calculations. \\
d) From the $1/N$-expansion at order $1/N^2$.
}
\end{table}

\newpage

\section*{Figures}

\renewcommand{\labelenumi}{Fig. \arabic{enumi}}
\begin{enumerate}
\item  
The evolution of $u'_k(\rht)$ as $k$ is lowered from $\Lx$ to zero.
The initial conditions (bare couplings)
have been chosen such that the scaling solution is approached
before the system evolves towards the symmetric phase
with $u'_k(0) > 0$.
$N=1$.
\item  
The evolution of $\kx$, $\lx$ and $\eta$
for the solution of fig. 1.
\end{enumerate}

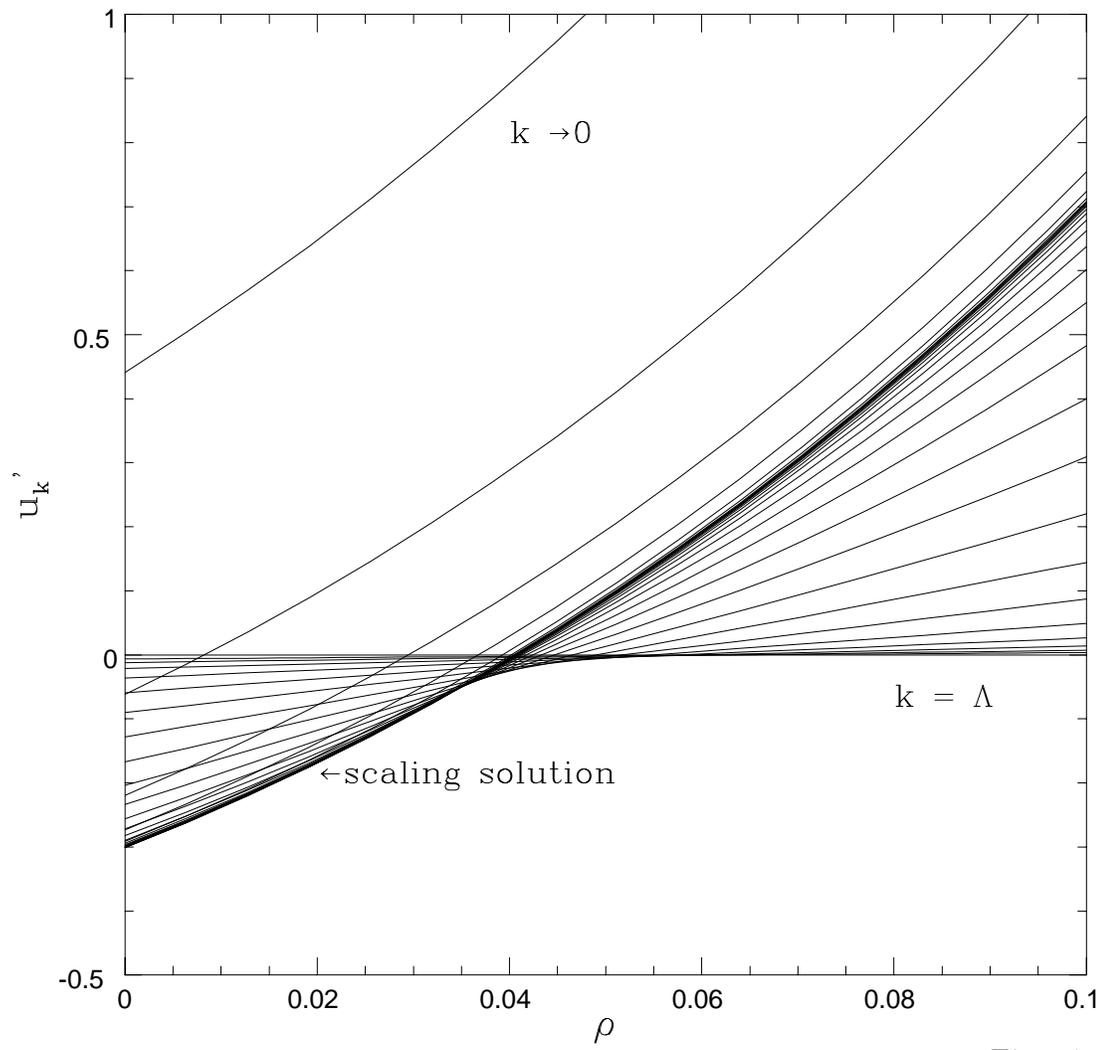
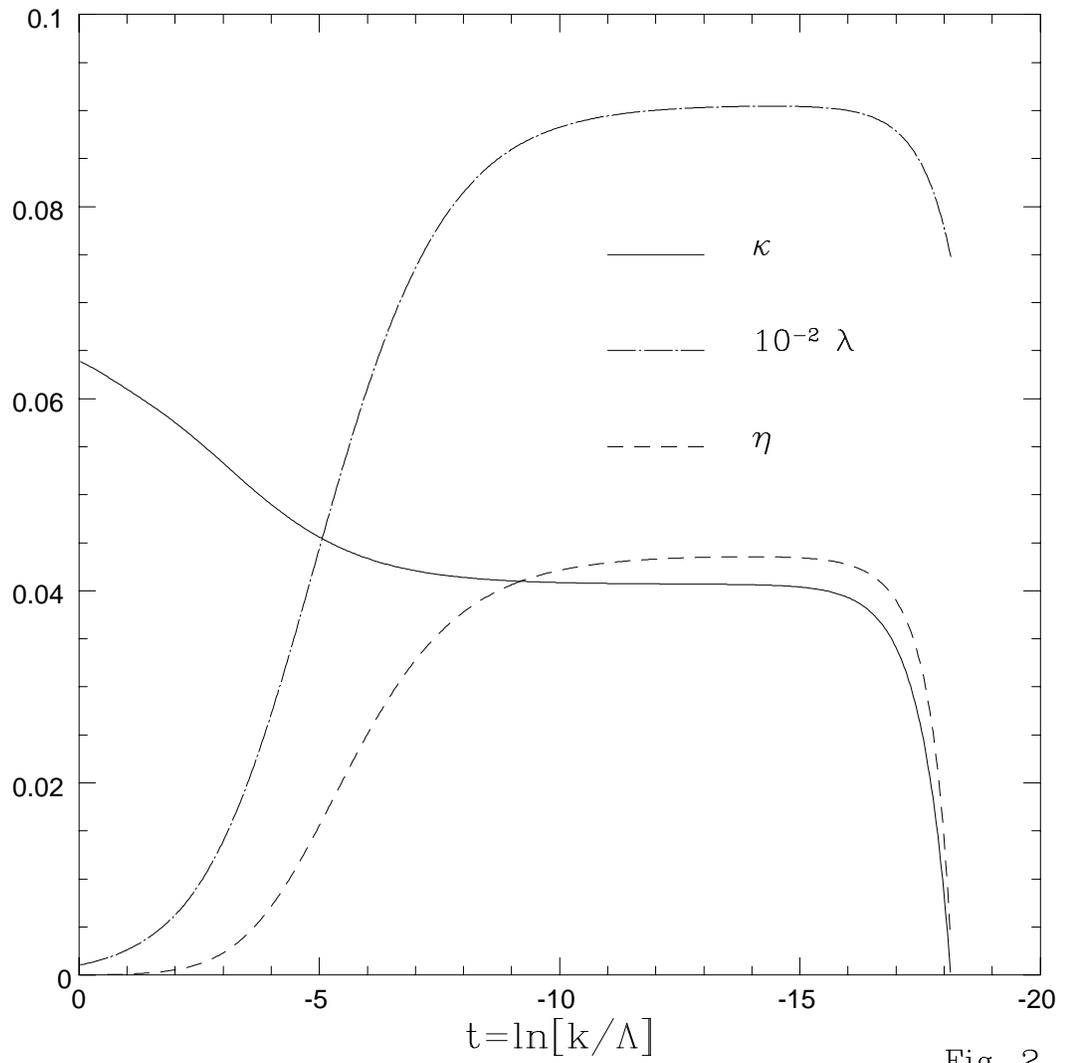
\end{document}